\documentclass[sigconf]{acmart}

\usepackage{booktabs}
\usepackage{xspace}
\usepackage{amssymb}
\usepackage{listings}
\usepackage{tikz}
\usepackage{amsmath}
\usepackage{algorithm}
\usepackage[noend]{algpseudocode}
\usepackage{pifont}
\usepackage{tabularx,booktabs}
\usepackage{flushend}
\usepackage{amsmath}
\usepackage{caption}
\usepackage{subcaption}
\PassOptionsToPackage{hyphens}{url}

\usepackage{enumitem}
\usepackage[T1]{fontenc}
\usepackage{hyperref}
\usepackage{multirow}
\usepackage{pgfplots}
\usepackage{soul}
\newcommand{\cmark}{\ding{51}}
\newcommand{\xmark}{\ding{55}}

\newcommand{\roundbox}[1]{
\begin{center}
	\begin{tikzpicture}
		\node[draw=black, rectangle, rounded corners](box){
    	\begin{minipage}{0.85\columnwidth}
        	#1
    	\end{minipage}
		};
	\end{tikzpicture}
\end{center}
}

\newcommand{\PA}[1]{\textbf{P#1}\xspace} 
\newcommand{\Q}[1]{\textbf{Q#1}\xspace} 
\newcommand{\RQ}[1]{\textbf{RQ#1}\xspace} 
\newcommand{\RQO}[2]{\textit{RQ#1-#2}\xspace} 

\definecolor{darkred}{rgb}{0.75,0,0}
\definecolor{darkblue}{rgb}{0,0,0.75}
\definecolor{darkgreen}{rgb}{0,0.75,0}

\newcommand{\quot}[1]{\textsf{\small{``#1''}}}

\newcommand{\hide}[1]{}
\newcommand{\empirical}[1]{#1}

\newcommand{\visuflow}{\textsc{Visuflow}\xspace}
\newcommand{\visuflowbr}{[\textsc{Visuflow}]\xspace}

\newcommand{\figref}[1]{Figure~\ref{#1}\xspace}
\newcommand{\secref}[1]{Section~\ref{#1}\xspace}
\newcommand{\tabref}[1]{Table~\ref{#1}\xspace}

\definecolor{pblue}{rgb}{0.13,0.13,1}
\definecolor{pgreen}{rgb}{0,0.5,0}
\definecolor{pred}{rgb}{0.9,0,0}
\definecolor{pgrey}{rgb}{0.46,0.45,0.48}
\algnewcommand\algorithmicswitch{\textbf{switch}}
\algnewcommand\algorithmiccase{\textbf{case}}
\algnewcommand\algorithmicassert{\texttt{assert}}
\algnewcommand\Assert[1]{\State \algorithmicassert(#1)}%
\algdef{SE}[SWITCH]{Switch}{EndSwitch}[1]{\algorithmicswitch\ #1\ \algorithmicdo}{\algorithmicend\ \algorithmicswitch}%
\algdef{SE}[CASE]{Case}{EndCase}[1]{\algorithmiccase\ #1}{\algorithmicend\ \algorithmiccase}%
\algtext*{EndSwitch}%
\algtext*{EndCase}%

\usetikzlibrary{fit,calc,trees,positioning,arrows,chains,shapes.geometric,decorations.pathreplacing,decorations.pathmorphing,shapes,matrix,shapes.symbols}
\tikzset{
>=stealth',
  punktchain/.style={
    rectangle, 
    rounded corners, 
    draw=black, very thick,
    text width=5em, 
    minimum height=3em, 
    text centered, 
    on chain},
  line/.style={draw, thick, <-},
  element/.style={
    tape,
    top color=white,
    bottom color=blue!50!black!60!,
    minimum width=5em,
    draw=blue!40!black!90, very thick,
    text width=5em, 
    minimum height=3.5em, 
    text centered, 
    on chain},
  every join/.style={->, thick,shorten >=1pt},
  tuborg/.style={decorate},
  tubnode/.style={midway, right=2pt},
}

\setcopyright{rightsretained}

\begin{document}

\title{Debugging Static Analysis}
\author{Lisa Nguyen Quang Do}
\affiliation{Fraunhofer IEM}
\email{lisa.nguyen@iem.fraunhofer.de}

\author{Stefan Kr{\"u}ger}
\affiliation{Paderborn University}
\email{stefan.krueger@upb.de}

\author{Patrick Hill}
\affiliation{Paderborn University}
\email{pahill@campus.uni-paderborn.de}

\author{Karim Ali}
\affiliation{University of Alberta}
\email{karim.ali@ualberta.ca}

\author{Eric Bodden}
\affiliation{\mbox{Paderborn University \& Fraunhofer IEM}}
\email{eric.bodden@upb.de}

\begin{abstract}

To detect and fix bugs and security vulnerabilities, software companies use static analysis as part of the development process. However, static analysis code itself is also prone to bugs. To ensure a consistent level of precision, as analyzed programs grow more complex, a static analysis has to handle more code constructs, frameworks, and libraries that the programs use. While more complex analyses are written and used in production systems every day, the cost of debugging and fixing them also increases tremendously.

To better understand the difficulties of debugging static analyses, we surveyed \empirical{115} static analysis writers. From their responses, we extracted the core requirements to build a debugger for static analysis, which revolve around two main issues: (1) abstracting from two code bases at the same time (the analysis code and the analyzed code) and (2) tracking the analysis internal state throughout both code bases. Most current debugging tools that our survey participants use lack the capabilities to address both issues.

Focusing on those requirements, we introduce \visuflow, a debugging environment for static data-flow analysis that is integrated in the Eclipse development environment. \visuflow features graph visualizations that enable users to view the state of a data-flow analysis and its intermediate results at any time. Special breakpoints in \visuflow help users step through the analysis code and the analyzed simultaneously. To evaluate the usefulness of \visuflow, we have conducted a user study on \empirical{20} static analysis writers. Using \visuflow helped our sample of analysis writers identify \empirical{25\%} and fix \empirical{50\%} more errors in the analysis code compared to using the standard Eclipse debugging environment.
\end{abstract}



\maketitle

\section{Introduction}

Software is getting more complex, with new features added every day to already-existing large code bases. To avoid coding errors, bugs, and security vulnerabilities, companies use various automated tools such as Google Tricorder~\cite{tricorder}, Facebook Infer~\cite{infer}, or HP Fortify~\cite{fortify}. One approach that is commonly used in these tools is static analysis, a method of automatically reasoning about the runtime behaviour of a program without running it.

As more complex software is produced, more complex analyses are also written to efficiently support bug finding. Prior static-analysis research has yielded many novel algorithms~\cite{ifds, incremental, prioritization}, analyses~\cite{aside, cheetah}, and analysis tools~\cite{tooling, findbugs} to better support code developers. However, standard debugging tools~\cite{eclipse, intellij, gdb} are often ill-suited to help analysis writers debug their own analyses. Writing an analysis is often more complex than writing general application code, because it requires thorough knowledge of both the analysis code and the analyzed code. Such knowledge enables the analysis writer to handle specific corner cases, while also ensuring soundness and precision. Precisely understanding what an analysis does is hard and time consuming, making the development of new analyses cumbersome in academia and industry.

In this paper, we investigate the need for better debugging tools for static analysis code through a large-scale survey of \empirical{115} static analysis writers. The survey aims to identify (1)~common types of static analysis, (2)~common bugs in static analysis code compared to application code, (3)~popular debugging tools used for static analysis code and application code, (4)~the limitations of those tools with respect to debugging static analysis code, and (5)~desirable features for a static analysis debugger.

Based on the debugging features that we have identified in the survey, we present \visuflow, an Eclipse-based debugging environment for data-flow analysis written on top of Soot~\cite{soot}. \visuflow helps analysis writers better visualize and understand their analysis code while debugging it. A focused user study with \empirical{20} participants shows that the debugging features of \visuflow help analysis writers identify \empirical{25\%} and fix \empirical{50\%} more errors in analysis code compared to using the standard Eclipse debugging environment. The participants found the debugging features in \visuflow more useful than their own coding environment for debugging static analysis. 

In summary, this paper makes the following contributions:
\begin{itemize}
\item Through a comprehensive survey, we motivate the need for better tools to debug static analyses, and identify desirable features such tooling should provide.
\item We present \visuflow, an Eclipse-based debugging environment for static data-flow analysis.
\item Through a user study, we evaluate the usefulness of \visuflow for debugging static analysis. We also determine which of the desirable features that we extracted from the survey are, in fact, useful for debugging static analysis.
\end{itemize}

\visuflow is available online, along with a video demo, the anonymized survey answers, and the results of the user study~\cite{visuflow}.
\section{Survey}
\label{sec:survey}

We conducted a large-scale survey of \empirical{115} static-analysis writers to understand the differences between debugging static analysis code and general application code. Our goal is to provide better support for the former through answering the following research questions:
\begin{enumerate}[label=\RQ{\arabic*:}]
\item Which types of static analysis are most commonly written?
\item Do analysis writers think that static analysis is harder/easier to debug than application code, and why?
\item Which errors are most frequently debugged in static analysis and application code?
\item Which tools do analysis writers use to support debugging of static analysis and application code?
\item What are the limitations of those tools?
\end{enumerate}

\subsection{Survey Design}
The survey contains \empirical{32} questions that we refer to as \Q{1}-\Q{32}, in the order in which they were presented to participants. The survey questions and anonymized answers are available online~\cite{visuflow}. We group the survey questions into the following \empirical{8} sections:

\begin{enumerate}
	\item \textbf{Participant information:} Through multiple-choice questions, we asked participants how long they have been writing static analysis (\Q{3}), for which languages (\Q{4}), and which branches (\Q{6}) and frameworks (\Q{9}) of static analysis they have experience with.
	
	\item \textbf{Debugging static analysis compared to application code:} \Q{11} asks participants which type of code is easier to debug on a scale from \empirical{1} (application code) to \empirical{10} (static analysis). \Q{12} asks them why in free text.
	
	\item \textbf{Debugging static analysis:} \Q{13} asks participants how long they spend on writing static analysis compared to debugging it on a scale from \empirical{0} (100\% coding, 0\% debugging) to \empirical{10} (0\% coding, 100\% debugging). In free text, \Q{15} asks for the typical causes of bugs they find in analysis code.

	\item \textbf{Tools for debugging static analysis:} In free text, we asked participants which features of their coding environments they like (\Q{17}), dislike (\Q{18}), and would like to have (\Q{19}) when debugging static analysis.

	\item \textbf{Debugging application code:} \Q{20} and \Q{21} are similar to \Q{13} and \Q{15}, but specific to application code.

	\item \textbf{Tools for debugging application code:} \Q{23}-\Q{25} are similar to \Q{17}-\Q{19}, applied to application code.

	\item \textbf{Specific debugging features:}  \Q{26} asks participants to rate the importance of some debugging features on the following scale: Not important - Neutral - Important - Very important - Not applicable.

	\item \textbf{Coding environment:} 
In a closed-choice question, \Q{28} asks participants if they primarily code using a text editor (e.g., Vim, Emacs) or an IDE (e.g., Eclipse, IntelliJ). \Q{29} asks them in free text which specific software they use.
\end{enumerate}

First, we sent a pilot survey to \empirical{10} participants. Based on the feedback, we refined the questions as shown above.

\subsection{Result Extraction}
We manually classified the answers to the free-text questions using an open card sort~\cite{cardsort}. Two authors classified the answers into various categories, which were derived during the classification process. Responses that do not answer the question were classified in an ``Others'' category (e.g., \quot{n/a}).

\begin{figure}[t]
\begin{tikzpicture}[trim axis right, trim axis left]
\begin{axis}[
    symbolic x coords={MIN, MIN2,\Q{1},\Q{3},\Q{4},\Q{6},\Q{9},\Q{11},\Q{12},\Q{13},\Q{15},\Q{17},\Q{18},\Q{19},\Q{20},\Q{21},\Q{23},\Q{24},\Q{25},\Q{26},\Q{28},\Q{29}, MAX2,MAX},
    xtick=data,
    bar width=9,
    x tick label style={rotate=90,anchor=east},
    ybar stacked,
    ytick={0, 20, 40, 60, 80, 100, 120},
    width=9cm,
    height=4.5cm,
    major x tick style = transparent,
    xmin=MIN,
    xmax=MAX  ]
    \addplot+[ybar,fill=lightgray,draw=black] plot coordinates {
        (\Q{1},115)        (\Q{3},112)        (\Q{4},106)        (\Q{6},106)
        (\Q{9},74)        (\Q{11},91)        (\Q{12},83)        (\Q{13},100)
        (\Q{15},48)        (\Q{17},71)       
        (\Q{18},44)        (\Q{19},38)        (\Q{20},83)        (\Q{21},32)        
        (\Q{23},54)        (\Q{24},28)        (\Q{25},17)        
        (\Q{26},75)        (\Q{28},75)        (\Q{29},75)        
    };
    \addplot+[ybar,fill=white,draw=black] plot coordinates {
        (\Q{1},0)        (\Q{3},0)        (\Q{4},0)        (\Q{6},0)
        (\Q{9},0)        (\Q{11},12)        (\Q{12},20)        (\Q{13},0)
        (\Q{15},19)        (\Q{17},17)       
        (\Q{18},44)        (\Q{19},17)        (\Q{20},0)        (\Q{21},14)        
        (\Q{23},22)        (\Q{24},48)        (\Q{25},15)        
        (\Q{26},0)        (\Q{28},0)        (\Q{29},0)        
    };
\end{axis}
\end{tikzpicture}
\caption{Number of anwsers that could (gray) or could not (white) be classified per question.}
\label{fig:survey_response_rate}
\end{figure}
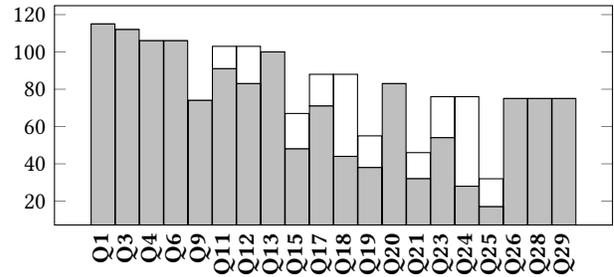

To verify the validity of our classification, another author, who had not been part of the classification phase nor seen the answers before, sorted the answers in the categories derived during the first classification. We then compared the agreement between the two classifications.
Since one answer could match multiple categories (e.g., ``I use breakpoints and stepping.'' matched the categories ``Breakpoint'' and ``Stepping''), we calculated a percent agreement for each category of each question. The average percent agreement over all categories for all questions is \empirical{96.3\%} (\empirical{median = 98\%}, \empirical{min = 65.2\%}, \empirical{max = 100\%}, standard deviation \empirical{$\sigma$ = 0.05}). Because of the imbalance in the distribution of the answers, we ran into a paradox of inter-rater agreement~\cite{kappaparadox}, making the Cohen's Kappa~\cite{kappa} an unreliable statistic for this survey (average~\empirical{$\kappa = 0.66$}, median~\empirical{$\kappa = 0.7$}, \empirical{$min = -0.08$}, \empirical{$max = 1$}, \empirical{$\sigma = 0.33$}).

Due to optional questions and participants who did not finish the survey, some questions received fewer answers than others. \figref{fig:survey_response_rate} reports the number of classified (gray) and unclassified (white) answers per question. In the following sections, the percentages reported for each question are based on the number of classified answers for the particular question and not on all~\empirical{115} answers. Participants could choose multiple answers to the same multiple-choice question, and an answer to a free-text question could match multiple categories. Therefore, the percentages for each question may add up to more than 100\%. 

\subsection{Participants}
\label{subsec:participants}
We contacted \empirical{450} researchers from the authors of static analysis papers published between 2014 and 2016 at the following conferences and their co-located workshops: ICSE, FSE, ASE, OOPSLA, ECOOP, PLDI, POPL, SAS. We received responses from \empirical{115} researchers,
~\empirical{85.2}\% from academia and~\empirical{15.7}\% from industry (\Q{1}).

Most participants are experienced static analysis writers. Approximately~\empirical{31.3}\% of the participants have 2--5~years of experience writing static analysis, \empirical{22.3}\% have 5--10~years of experience, \empirical{26.8}\% have more than 10~years of experience, and only~\empirical{9.8}\% have less than 2~years of experience (\Q{3}). 

\subsection{Results}

\begin{figure}
\begin{tikzpicture}[trim axis right, trim axis left]
\begin{axis}[
    symbolic x coords={MIN, 1, 2, 3, 4, 5, 6, 7, 8, 9, 10, MAX},
    xtick=data,
    bar width=10,
    ytick={0, 5, 10, 15, 20, 25},
    yticklabel=\pgfmathprintnumber{\tick}\,\%,
    width=8cm,
    height=4cm,
    major x tick style = transparent,
    ymin=0,
    ymax=25,
    xmin=MIN,
    xmax=MAX  ]
    \addplot [ybar,fill=lightgray,draw=black] coordinates { (1,11.6) (2,10.7) (3,19.4) (4,8.7) (5,22.3) (6,5.8) (7,3.8) (8,2.9) (9,1.9) (10,1) };
\end{axis}
\end{tikzpicture}
\caption{Ranking the difficulty of debugging static analysis code compared to application code on a scale from 1 (static analysis is harder) to 10 (application code is harder). (\Q{11})}
\label{fig:survey_vs}
\end{figure}
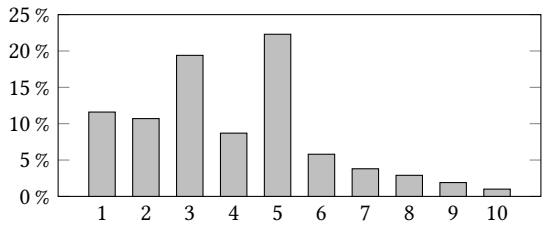

\subsubsection{\RQ{1}: Which types of static analysis are most commonly written?}

We asked participants which programming language they analyzed the most (\Q{4}), and received 3 main answers: Java (\empirical{62.3}\%), C/C++ (\empirical{59.4}\%), and JavaScript (\empirical{23.6}\%). There were~\empirical{34} other named languages, each analyzed by less than~\empirical{8}\% of the participants.

We also asked participants what branches of static analysis they wrote for (\Q{6}). Data-flow analysis is the most popular (\empirical{74.5}\%), followed by abstract interpretation (\empirical{65.1}\%), symbolic execution (\empirical{36.8}\%), and model checking (\empirical{21.7}\%). The remaining~\empirical{9} categories are each used by less than~\empirical{4\%} of the participants.

Finally, we asked participants about the most popular frameworks they used to write static analysis (\Q{9}). Soot~\cite{soot} is first (\empirical{55.4}\%), followed by WALA~\cite{wala} and LLVM~\cite{llvm} as second and third (\empirical{31.1}\% and \empirical{21.6}\%, respectively). The participants named~\empirical{32} other frameworks, each is used by less than~\empirical{10}\% of them.

\roundbox{
\RQO{1}{1}: Java is the most analyzed programming language. Data-flow analyses are the most common type of static analysis. Soot is the most popular analysis framework.
}

\subsubsection{\RQ{2}: Do analysis writers think that static analysis is harder/easier to debug than application code, and why?}
\label{subsec:rq2}

\Q{11} asks participants to rate how hard debugging static analysis is compared to debugging application code on a scale from 1 (static analysis is harder to debug) to 10 (application code is harder to debug). The average ranking is \empirical{4.0} (standard deviation \empirical{$\sigma = 2.1$}). \figref{fig:survey_vs} shows that~\empirical{50.5}\% of the participants find static analysis harder to debug than application code, \empirical{28.2}\% are neutral, and \empirical{9.5}\% think that application code is harder to debug. 
\Q{13} and \Q{20} confirm this result. Participants reported that they spent more time debugging a piece of static analysis code than writing it, and the contrary for a piece of application code. On average, participants estimate spending \empirical{46.8}\% of their time writing static analysis code compared to \empirical{57.5}\% writing application code. 
A $\mathcal{X}^2$ test of independence does not detect a significant correlation \empirical{($p > 0.05$)} between the rating of \Q{11} and the participant background (seniority, coding languages, editor type, or analysis frameworks).

\begin{table}[t]
\centering
\caption{Reasons why static analysis is harder to debug than application code (SA) and vice-versa (AC). EQ denotes the reasons why both are equally difficult to debug. (\Q{12})}
\label{fig:survey_vs_why}
\resizebox{\columnwidth}{!}{
\begin{tabular}{ l l r }
\toprule
\textbf{Harder} & \textbf{Reason} & \textbf{\%} \\
\midrule
\multirow{7}{*}{SA} & Abstracting two types of code & 15.67\% \\
& Greater variety of cases & 15.7\% \\ 
& More complex structure of static analysis tools & 6.0\%  \\ 
& Evaluating correctness is harder & 6.0\%  \\ 
& Soundness is harder to achieve & 3.6\%  \\ 
& Intermediate results are not directly accessible & 4.8\%  \\ 
& Static analysis is harder to debug & 3.6\%  \\ 
\midrule
\multirow{3}{*}{EQ} & Both are application code & 13.3\% \\
& They cannot be compared & 7.2\% \\
& No opinion & 3.6\% \\
\midrule
\multirow{3}{*}{AC} & Used to developing static analysis & 6.0\% \\
& Application code is more complex & 2.4\% \\
\bottomrule
\end{tabular}}
\end{table}

\tabref{fig:survey_vs_why} classifies the reasons that participants gave when asked why they found one type of code harder to debug than the other (\Q{12}). 
Due to space limitations, we only report the reasons mentioned by more than one participant. The main reason that participants find static analysis harder to debug is the complexity of handling two code bases (i.e., the analysis code and the analyzed code) at the same time: \quot{Static Analysis requires to switch between your analysis code and the Intermediate Representation which you actually analyse}. This complexity creates more corner cases that the analysis writer has to handle. 
Another reason is that correctness is harder to define for a static analysis. To quote a participant: \quot{`correct' is better defined [in application code]}.
The final reason is that analysis intermediate results are not directly verifiable, as opposed to the output of application code that can be directly validated: \quot{Static analysis code usually deals with massive amounts of data. [...] It is harder to see where a certain state is computed, or even worse, why it is not computed.}

Participants who find static analysis and application code equally hard to debug have two main arguments. First, both are application code: \quot{a static analyzer is an application, albeit a sophisticated one}. Second, they are so different that they cannot be compared: \quot{These two difficulties are qualitatively different and hence incomparable.} 

Participants who find application code more difficult to debug argue that it is more complex than static analysis code. Depending on the degree of complexity of the application and the analysis, application code may contain a huge number of corner cases that the application writer has to comprehend: \quot{Static analysis code usually includes very limited number of possible cases.} Some participants also wrote that the reason why they find application code harder to debug is that they are used to developing static analysis.

\roundbox{
\RQO{2}{1}: \empirical{5.3}$\times$ more participants found static analysis harder to debug than application code. This is due to three main reasons: handling two code bases simultaneously, correctness requirements for static analysis, and the lack of support for debugging static analysis.
}

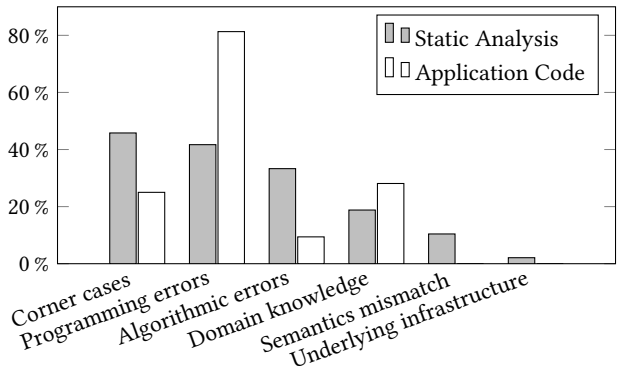
\begin{figure}[t]
\begin{tikzpicture}[trim axis right, trim axis left]
    \begin{axis}[
    symbolic x coords={MIN,Corner cases,Programming errors,Algorithmic errors,Domain knowledge,Semantics mismatch,Underlying infrastructure,MAX},
    major x tick style = transparent,
    xtick=data,
    bar width=10,
    x tick label style={align=center,rotate=20,anchor=east},
    ybar=2*\pgflinewidth,    
    ytick={0, 20, 40, 60, 80, 100},
    yticklabel=\pgfmathprintnumber{\tick}\,\%,
    width=9cm,
    height=5cm,
    xmin=MIN,
    xmax=MAX,
    scaled y ticks = false,
    ymin=0,
    ymax=90,
    legend cell align=left,
    legend pos=north east
    ]
    \addplot[style={draw=black,fill=lightgray,mark=none}]
            coordinates {(Corner cases, 45.8) (Programming errors,41.7) (Algorithmic errors,33.3) (Domain knowledge,18.8) (Semantics mismatch,10.42) (Underlying infrastructure,2.1)};
    \addplot[style={draw=black,fill=white,mark=none}]
            coordinates {(Corner cases, 25.0) (Programming errors,81.3) (Algorithmic errors,9.4) (Domain knowledge,28.1) (Semantics mismatch,0) (Underlying infrastructure,0)};
    \legend{\strut Static Analysis, \strut Application Code}
    \end{axis}
\end{tikzpicture}
\caption{The root causes of errors found when debugging static analysis and application code. (\Q{15} and \Q{21})}
\label{fig:survey_bug_causes}
\end{figure}

\subsubsection{\RQ{3}: Which errors are most frequently debugged in static analysis and application code?}

We asked the participants for the typical root causes of errors they find when debugging static analysis (\Q{15}) and application code (\Q{21}). We categorize the results in the six categories shown in \figref{fig:survey_bug_causes}. 
When debugging static analysis, the main cause of errors is handling \emph{corner cases}, which is \empirical{2$\times$} more prevalent than when writing application code. This category includes overlooked cases that the developer normally knows of (\quot{Forgot to consider the effect of certain, rare instructions}). The \emph{domain knowledge} category refers to code behaviour that the developer is unaware of (\quot{Unexpected values returned by an API}). 

\emph{Programming errors} occur \empirical{2$\times$} more often in application code than in static analysis code. This category includes implementation errors such as \quot{wrong conditions, wrong loops statements}. The category \emph{algorithmic errors} contains errors due to a wrong design decision in the program's algorithm such as a \quot{non-convergence} of the analysis (\Q{15}) or an issue with \quot{the algorithms I'm implementing} (\Q{21}). Participants debug such algorithmic errors \empirical{3.5$\times$} more often in static analysis than in application code.

\textit{Semantics mismatch} and \textit{underlying infrastructure} are specific to static analysis. The former refers to how the analysis interprets the analyzed code (e.g., \quot{The code does not take [into] account the abstract semantics correctly}). The latter is similar to \textit{domain knowledge}, but instead of the knowledge of the analyzed code, it is about the analysis framework (e.g., \quot{Can't load classes/methods successfully.}).

While bugs in application code are mainly due to programming errors, static analysis bugs are distributed over multiple categories. We attribute this to the heightened interest of analysis writers to produce correct analyses. Testing functional correctness typically requires validating input/output relationships. For static analysis, those relationships are always imperfect due to necessary approximations. Hence, it is very hard to define functional correctness for static analysis.
Moreover, handling two code bases is also the cause of analysis-specific errors: semantics mismatch and underlying infrastructure.
Because of the specific requirements of static analyses, the bugs that developers investigate in application code have different causes compared to analysis code. Therefore, there is a dire need to support debugging static analysis for the specific kind of errors that are of interest to analysis writers.

\roundbox{
\RQO{3}{1}: Programming errors are the main cause of bugs in application code. Static analysis writers additionally debug analyses for corner cases, algorithmic errors, semantics mismatch, and unhandled cases in the underlying analysis infrastructure.
}

\subsubsection{\RQ{4}: Which tools do analysis writers use to support debugging of static analysis and application code?}

In \Q{28} and \Q{29}, \empirical{56}\% of the participants answered that to write static analysis, they use an Integrated Development Environment (IDE) such as Eclipse~\cite{eclipse} (used by \empirical{28}\%) or IntelliJ~\cite{intellij} (\empirical{17.3}\%), while \empirical{42.7}\% use text editors such as Vim~\cite{vim} (\empirical{33.3}\%) or Emacs~\cite{emacs} (\empirical{21.3}\%). 
Each of the other \empirical{21}~tools is used by less than \empirical{10}\% of the participants.

\begin{table}[t]
\centering
\caption{Useful features for debugging static analysis (SA) and application code (AC) for IDE users (IDE) and text editor users (TE). (\Q{17} and \Q{23})}
\label{fig:survey_features_liked}
\resizebox{\columnwidth}{!}{
\begin{tabular}{ l c c c c }
\toprule
 & \textbf{SA/IDE} & \textbf{SA/TE} & \textbf{AC/IDE} & \textbf{AC/TE} \\
\midrule
Printing & \cmark & \cmark & \cmark & \cmark \\
Breakpoints & \cmark & \cmark & \cmark & \cmark \\
Debugging tools & \cmark & \cmark & \cmark & \cmark \\
Coding support & \cmark & \cmark & \cmark & \cmark \\
Variable inspection & \cmark & \cmark & \cmark & \cmark \\
Automated testing & \cmark & \cmark & \cmark & \cmark \\
Expression mode & \cmark & \cmark & \cmark & \cmark \\
\midrule
Memory tools &  & \cmark & \cmark & \cmark \\
\midrule
Graph visualizations &  & \cmark &  & \cmark \\
Stepping & \cmark &  & \cmark &  \\
Type checker & \cmark &  & \cmark &  \\
Hot-code replacement & \cmark &  & \cmark &  \\
\midrule
Visualizations &  &  &  & \cmark \\
Stack traces &  & \cmark &  &  \\
Drop frames &  &  & \cmark &  \\
Documentation & \cmark &  &  &  \\
\bottomrule
\end{tabular}}
\end{table}

We asked the participants about the most useful features of their coding environments when debugging static analysis (\Q{17}) and application code (\Q{23}). \tabref{fig:survey_features_liked} shows the features mentioned by more than one participant.
The most popular debugging feature is Breakpoints, used by \empirical{35.2}\% of participants when debugging application code and \empirical{28.2}\% for static analysis. Coding support (e.g., auto-completion) is appreciated by \empirical{29.6}\% when writing static analysis, and \empirical{20.4}\% for application code. Variable inspection is used by \empirical{27.8}\% of the application code writers and \empirical{19.7}\% of the analysis writers. Debugging tools (e.g., \quot{GDB/JDB}) are used by \empirical{20.4}\% when writing application code, and \empirical{16.9}\% for analysis code.
Printing intermediate results is used by \empirical{21.1}\% to debug analysis code, compared to \empirical{13.0}\% for application code. IDE users highlighted IDE-specific features such as type checkers, stepping, and hot-code replacement.

A $\mathcal{X}^2$ test of independence shows a strong correlation between the type of editor used (IDE or text editor) and the most useful features of the coding environment \empirical{($p = 0.01 \leq 0.05$)} for application code. The test does not find such a correlation for static analysis.

\roundbox{
\RQO{4}{1}: Analysis writers almost equally use IDEs and text editors. Regardless of the coding environment, analysis writers and application code writers use the same debugging features such as breakpoints, variable inspection, coding support, and printing intermediate results.
}

\subsubsection{\RQ{5}: What are the limitations of the existing debugging tools?}
\label{subsubsec:rq5}

\begin{table}[t]
\centering
\caption{Unsatisfactory features when debugging static analysis (SA) and application code (AC) for IDE users (IDE) and text editor users (TE). (\Q{18} and \Q{24})}
\label{fig:survey_features_disliked}
\resizebox{\columnwidth}{!}{
\begin{tabular}{ l c c c c }
\toprule
 & \textbf{SA/IDE} & \textbf{SA/TE} & \textbf{AC/IDE} & \textbf{AC/TE} \\
\midrule
Debugging tools & \xmark & \xmark & \xmark & \xmark \\
Immediate feedback & \xmark & \xmark & \xmark & \xmark \\
Coding support & \xmark & \xmark & \xmark & \xmark \\
\midrule
Multiple environments &  & \xmark & \xmark & \xmark \\
\midrule
Intermediate results & \xmark & \xmark &  &  \\
Handling data structures & \xmark &  & \xmark &  \\
Support for system setup &  &  & \xmark & \xmark \\
\midrule
Scalability & \xmark &  &  &  \\
Visualizations & \xmark &  &  &  \\
Conditional breakpoints & \xmark &  &  & \\
Memory tools &  &  & \xmark &  \\
Bad documentation &  &  &  & \xmark \\
\bottomrule
\end{tabular}}
\end{table}

\Q{18} and \Q{24} ask participants about the features of their coding environments they dislike when debugging static analysis and application code, respectively. The features mentioned by more than one participant are shown in \tabref{fig:survey_features_disliked}.
Debugging tools lack features to support debugging static analysis according to \empirical{29.5}\% of participants, compared to \empirical{25}\% when debugging application code. The lack of immediate feedback when a change is made to the code was noted by \empirical{11.4}\% of analysis writers and \empirical{17.9}\% of application code writers. Coding support for static analysis is disliked by \empirical{18.2}\% and \empirical{25}\% for application code.

To our surprise, two of the most disliked features---debugging tools and coding support---are also among the most used and appreciated. This suggests that although current tools are useful, users require more specific features to fully support their needs. For example, a participant wrote: \quot{While the IDE can show a path through [my] code for a symbolic execution run, it doesn't show analysis states along that path.} Therefore, debugging tools for static analysis could be improved by showing more of the intermediate results of the analysis. For application code, participants requested more support for handling different systems and environments. Participants complained about the \quot{manual work to setup complex build/test systems} and \quot{Dealing with an external dependency (e.g., communicating with a party that I cannot control)}).
Static analysis writers using an IDE find debugging tools not scalable, lack of visualizations of analysis constructs (e.g., \quot{It's mostly text-based}), and need special breakpoints (e.g., \quot{Missing an easy way to add a breakpoint when the analysis reaches a certain line in the input program (hence having to re-run an analysis)}).

\roundbox{
\RQO{5}{1}: Current static-analysis debugging tools lack important features such as showing intermediate results, providing clear visualizations of the analysis, and special breakpoints.
}

\begin{table}[t]
\centering
\caption{Requested features when debugging static analysis (SA) and application code (AC) for IDE users (IDE) and text editor users (TE). (\Q{19} and \Q{25})}
\label{fig:survey_features_wanted}
\resizebox{\columnwidth}{!}{
\begin{tabular}{ l c c c c }
\toprule
 & \textbf{SA/IDE} & \textbf{SA/TE} & \textbf{AC/IDE} & \textbf{AC/TE} \\
\midrule
Graph visualizations & \cmark & \cmark &  &  \\
Omniscient debugging & \cmark & \cmark &  &  \\
Visualizations & \cmark & \cmark &  &  \\
Hot-code replacement & \cmark &  & \cmark &  \\
Coding support &  &  & \cmark & \cmark \\
\midrule
Test generation &  & \cmark &  &  \\
Debugging tools &  & \cmark &  &  \\
Intermediate results & \cmark &  &  &  \\
Conditional breakpoints & \cmark &  &  &  \\
Handling data structures &  &  & \cmark &  \\
\bottomrule
\end{tabular}}
\end{table}

To identify which debugging features would best support static analysis writers, we asked the participants to suggest useful features for debugging static analysis (\Q{19}) and for debugging application code (\Q{25}). \tabref{fig:survey_features_wanted} shows the features that are mentioned more than once. The requested debugging features for application code and static analysis are quite different. To write application code, participants requested better hot-code replacement and coding support (e.g., \quot{better support to record complex data coming from external services}). For static analysis, \empirical{18.4}\% of participants asked for better visualizations of the analysis constructs, and \empirical{23.7}\% requested graph visualizations: \quot{Easier way to inspect `intermediate' result of an analysis, easier way to produce state graphs and inspect them with tools.} Omniscient debugging was requested by \empirical{13.2}\% of participants to help show the intermediate results of the analysis: \quot{Stepping backwards in the execution of a program}. Participants also requested better test generation tools and special breakpoints (\RQO{5}{1}).

A $\mathcal{X}^2$ test on the features of \tabref{fig:survey_features_wanted} shows a correlation between the type of code written (static analysis or application code) and the features requested by participants \empirical{($p = 0.04 \leq 0.05$)}. The same test does not show correlations between the type of code and the features liked and disliked by participants, with $p$-values of \empirical{0.97} and \empirical{0.69}, respectively.
In addition, the test shows strong correlations between the type of editor used and the requested features for static analysis \empirical{($p = 0.02$)} and application code \empirical{($p = 0.04$)}. Such correlations could not be shown for features that participants liked or disliked, suggesting that the tools used to debug application code and static analysis contain features that all types of users equally like and dislike. Regardless of the editor, the requested features for writing static analysis and application code are quite different.

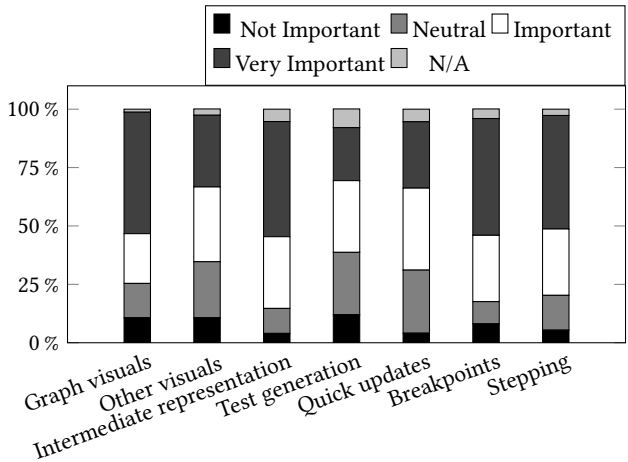
\begin{figure}[t]
\begin{tikzpicture}[trim axis right, trim axis left]
    \begin{axis}[
    symbolic x coords={MIN,Graph visuals,Other visuals,Intermediate representation,Test generation,Quick updates,Breakpoints, Stepping,MAX},
    major x tick style = transparent,
    xtick=data,
    bar width=10,
    x tick label style={align=center,rotate=20,anchor=east,yshift=-3mm,xshift=2mm},
    ybar stacked,    
    yticklabel=\pgfmathprintnumber{\tick}\,\%,
    ytick={0, 25, 50, 75, 100},
    width=9cm,
    height=5cm,
    xmin=MIN,
    xmax=MAX,
    ymax=110,
    scaled y ticks = false,
    ymin=0,
    legend style={at={(1,1)},anchor= south east, legend columns=3}
    ]
    
    \addplot+[style={draw=black,fill=black,mark=none}]
            coordinates {(Graph visuals, 10.7) (Other visuals,10.7) (Intermediate representation,4) (Test generation,12) (Quick updates,4.1) (Breakpoints,8.1) (Stepping,5.4)};
		\addplot+[style={draw=black,fill=gray,mark=none}]
          coordinates {(Graph visuals, 14.7) (Other visuals,24) (Intermediate representation,10.7) (Test generation,26.7) (Quick updates,27) (Breakpoints,9.5) (Stepping,14.9)};
	  \addplot+[style={draw=black,fill=white,mark=none}]
            coordinates {(Graph visuals, 21.3) (Other visuals,32) (Intermediate representation,30.7) (Test generation,30.7) (Quick updates,35.1) (Breakpoints,28.4) (Stepping,28.4)};
		\addplot+[style={draw=black,fill=darkgray,mark=none}]
            coordinates {(Graph visuals, 52) (Other visuals,30.7) (Intermediate representation,49.3) (Test generation,22.7) (Quick updates,28.4) (Breakpoints,50) (Stepping,48.6)};
		\addplot+[style={draw=black,fill=lightgray,mark=none}]
            coordinates {(Graph visuals, 1.3) (Other visuals,2.7) (Intermediate representation,5.3) (Test generation,8) (Quick updates,5.4) (Breakpoints,4.1) (Stepping,2.7)};
    \legend{\strut Not Important, \strut Neutral, \strut Important, \strut Very Important, \strut N/A}
    \end{axis}
\end{tikzpicture}
\caption{Ranking the importance of features for debugging static analysis. (\Q{26})}
\label{fig:survey_features_importance}
\end{figure}

In \Q{26}, the participants evaluate how important such features are to them. \figref{fig:survey_features_importance} shows that graph visuals, access to the intermediate representation and intermediate results count as very important features, along with breakpoints and stepping functionalities that consider both the analysis code and the analyzed code. Participants find other types of visuals, better test generation, and quick updates less important.

\roundbox{
\RQO{5}{2}: Participants requested quite different features to debug static analysis compared to application code. The most important features for debugging static analysis are: graph visualizations, access to analysis intermediate results, and conditional breakpoints.
}

\subsection{Discussion}
\label{subsec:intermediate_discussion}

This survey shows that writing static analysis entails specific requirements on the writer. Handling two code bases and defining soundness makes static analysis harder to debug than general application code (\RQO{2}{1}). Those requirements cause different types of bugs to be of interest to static analysis writers when debugging. In addition to programming errors, corner cases and algorithmic errors are specific to static analysis code (\RQO{3}{1}). To debug their code, analysis writers mainly use the traditional debugging features included in their coding environments such as breakpoints and variable inspection (\RQO{4}{1}). While those tools are helpful, they are not sufficient to fully support debugging static analysis. Debugging features such as simple breakpoints fall short and force analysis writers to handle parts of the debugging process manually (\RQO{5}{1}).

\tabref{fig:survey_features_liked} shows that the debugging tools that analysis writers currently use are adapted for more general application code. \tabref{fig:survey_features_wanted} shows that the features needed to debug analysis code are quite different from the features needed to debug application code.
To improve the debug process of static analysis, we identify and prioritize the following features: graph visualizations, access to the analysis intermediate representation and results, and breakpoints that can be controlled on the analysis and the analyzed code (\RQO{5}{2}).

In the remainder of this paper, we present a tool that supports features for debugging static analysis, based on the observations from our survey. We also conduct a user study to test how useful those features are for debugging static analysis, and discuss its results. To reach the largest possible target audience, we focus on the most popular use case: providing debugging support in the IDE (\RQO{4}{1}) for data-flow analysis of Java programs using the Soot analysis framework (\RQO{1}{1}).

\section{\visuflow: Visual Support for Debugging Data-Flow Analysis}
\label{sec:implementation}

\begin{figure*}[t]
    \centering
    \includegraphics[width=\textwidth]{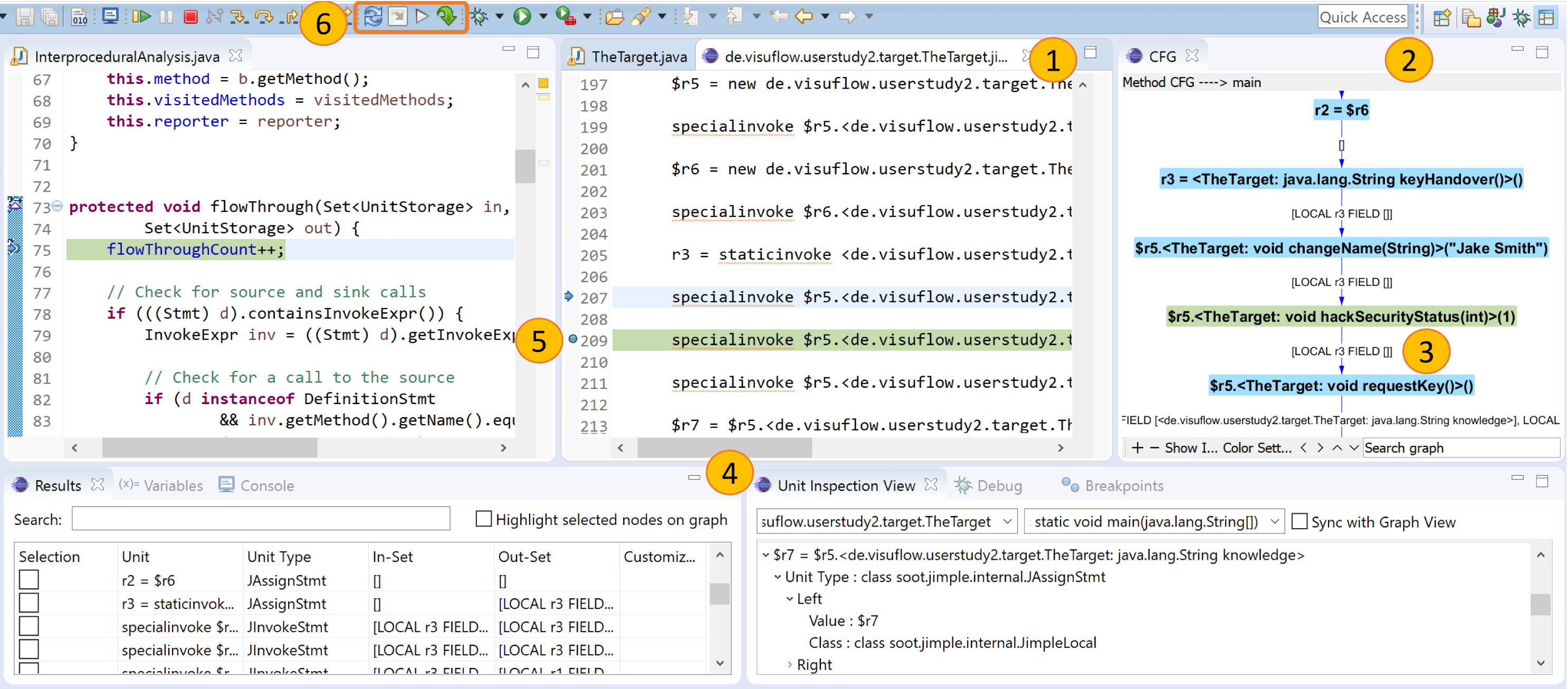}
    \caption{The Graphical User Interface of \visuflow. Features 1--6 are detailed in \secref{sec:implementation}.}
    \label{fig:ui}
\end{figure*}

We present \visuflow, an Eclipse-based debugging environment designed for static data-flow analyses written on top of the Soot analysis framework~\cite{soot}. \figref{fig:ui} presents the main Graphical User Interface (GUI) elements in \visuflow. These elements, described below and highlighted with the corresponding numbers in the figure, implement several features identified in \secref{subsubsec:rq5}. A video demo of the GUI is available online~\cite{visuflow}.

\begin{enumerate}[label=\textbf{\arabic*.},leftmargin=\parindent]

\item \textbf{Intermediate representation:}
Before running an analysis, Soot translates the program into a Jimple intermediate representation~\cite{soot} that is simpler to analyze. The Jimple code can be very different from the original source code. Therefore, \visuflow displays the Jimple code in its own read-only \emph{Jimple View}.

\item \textbf{Graph visuals:} 
To help the user better visualize the structure of the analyzed code, \visuflow has a \emph{Graph View} that displays the call graph and control-flow graphs (CFGs) of the analyzed code.
The user can customize the color of specific graph nodes or certain types of nodes (e.g., assignment statements). 

\item \textbf{Access to intermediate results:} 
In the graph, intermediate analysis results are edge labels. 
The user can then determine with a quick glance where a particular data-flow fact is generated, killed, or transferred, instead of debugging the analysis statement after statement to manually inspect intermediate results. 
The view is populated while debugging the analysis and generating its results. Intermediate results are displayed in the \emph{Results View}.

\item \textbf{Visuals:} 
The default layout of \visuflow has the analysis code and the analyzed code side-by-side, enabling the user to see both code bases and the graph in the same view, instead of switching from one tab to the other. The \emph{Results View} provides a compact summary of the \emph{Graph View}. The \emph{Unit Inspection View} shows a list of the program statements, allowing users to inspect the details of what a statement comprises (i.e., type of the statement and types of its components). This feature is useful to novice Soot users who might have little or no knowledge of Jimple, but need to handle particular types of statements.

\item \textbf{Debugger:}
Traditional debuggers allow static-analysis writers to set breakpoints only in the analysis code. To debug an analysis for a specific statement of the analyzed code, users have to use conditional breakpoints that suspend the analysis only for that statement. This is difficult when users do not know which statements they should inspect (i.e., when they need to step through the analyzed code). Using \visuflow, users can set breakpoints in both the analysis code and the \emph{Jimple View}. \visuflow provides separate stepping functionalities to help users debug through both code bases at the same time. 
\visuflow retains the basic Eclipse \emph{Variable Inspection View} and \emph{Stack Frame View}. 

\item \textbf{Linking analysis code and analyzed code}: 
	\visuflow introduces a new type of Eclipse project dedicated to static analysis. This project type provides code stubs for both the analysis code and the analyzed code, and links them in the background. This allows \visuflow to identify where to extract the necessary information when displaying analysis results on the graph representing the analyzed code. It is also possible to link an existing analysis to an existing application project.
	
\item \textbf{View synchronization}: 
The \visuflow views can be synchronized around a particular statement that is being observed by the user. This avoids confusion when browsing through different views, as observed during our pilot user study.
\end{enumerate}

\section{User Study}
\label{sec:userstudy}

We conducted a user study to evaluate how useful \visuflow is as a debugging tool for static analysis through answering the following research questions:
\begin{enumerate}[label=\RQ{\arabic*:}]
\setcounter{enumi}{5}
\item Which features of \visuflow are used the most?
\item Does \visuflow help identify and fix more errors compared to the standard debugging environment in Eclipse?
\item Does \visuflow help understand and debug analyses better than other debugging environments?
\end{enumerate}

\subsection{Setup}
\label{subsec:study_setup}
We evaluate how users interact with \visuflow compared to the standard Eclipse debugging environment~\cite{eclipse} (hereafter, referred to as Eclipse).
Each participant performed two tasks: debugging a static analysis once with \visuflow and once with Eclipse. In the latter case, participants had access to various Eclipse debugging functionalities such as breakpoints, stepping, the variables view, and the stack frame view. To ensure a fair comparison with \visuflow, we provided participants with the Jimple intermediate representation of the analyzed code when they used Eclipse. 

The two subject analyses we used are hand-crafted taint analyses that contain three errors each. Running either analysis on given programs does not compute the correct results. For each task, a participant had \empirical{20}~minutes to identify and fix as many errors as possible in the analysis code. 
Half of the participants performed their first task with \visuflow, and the other half with Eclipse. Both groups switched tools for the second task. Before each task, participants were given time to work on a demo analysis to familiarize themselves with the tools. They then performed their task on a different analysis and analyzed code.
We calibrated the difficulty of the tasks through a pilot study that we conducted on \empirical{6} participants.

During the two tasks, we measured the number of errors that participants identified and fixed. We also logged how long the mouse focus was for each view of the coding environment to evaluate the time spent using each view.
Afterwards, participants were asked to fill a comparative questionnaire of the two debugging environments, followed by a short discussion of their impressions about the tools. The full-text answers presented in \secref{subsec:study_usefulness} were categorized by two authors in an open card sort~\cite{cardsort}. Because each answer could be categorized in multiple categories, we calculated a Cohen's Kappa for each category of each question. The average Kappa score over all questions and categories is~\empirical{$\kappa$ = 0.98} (\empirical{median = 1}, \empirical{min = 0.66}, \empirical{max = 1}, standard deviation~\empirical{$\sigma$ = 0.07}), which is above the 0.81 threshold, indicating an almost perfect agreement~\cite{kappa}. The user study results are available online~\cite{visuflow}.

\subsection{Participants}
A total of \empirical{20} people participated in our user study (referred to as \PA{1}--\PA{20}). Study participants are of diverse backgrounds: researchers in academia, researchers in industry, and students. \empirical{Eleven} participants have less than a year experience writing static analysis, \empirical{6} have 1--5 years of experience, and \empirical{3} have more than 5 years of experience.

Participants rated their familiarity of data-flow analysis, Eclipse, and Soot on a scale from 0 (novice) to 10 (expert). The average score is \empirical{5.7} (min: \empirical{1}, max: \empirical{9}) for data-flow analysis, \empirical{5.9} (min: \empirical{2}, max: \empirical{8}) for Eclipse, and \empirical{3.3} (min: \empirical{0}, max: \empirical{7}) for Soot.
We thus gathered a variety of both novice and expert users in data-flow analysis and Eclipse. However, expert Soot users are rare.
Only \PA{7} and \PA{8} participated in the survey whose results motivate the main features of \visuflow, ensuring the impartial evaluation of the remaining \empirical{18} participants.

\subsection{Results}

\subsubsection{\RQ{6}: Which features of \visuflow are used the most?}
\begin{table}[t]
\centering
\caption{Main features of \visuflow and Eclipse that participants used, and the average time spent using each feature.}
\label{fig:study_features_used}
\resizebox{\columnwidth}{!}{
\begin{tabular}{ l r r r r }
\toprule
 & \multicolumn{2}{ c }{\textbf{\visuflow}} & \multicolumn{2}{ c }{\textbf{Eclipse}} \\
 \cmidrule(lr){2-3} \cmidrule(lr){4-5}
 & \textbf{\#users} & \textbf{Time (s)} & \textbf{\#users} & \textbf{Time (s)} \\
\midrule
Java Editor & 14 & 486 & 14 & 653 \\
Graph View & 14 & 201 & n/a & n/a \\
Jimple View & 11 & 58 & 12 & 60 \\
Breakpoints / Stepping & 11 & 174 & 11 & 313 \\
Variable Inspection & 3 & 78 & 8 & 67 \\
Results View & 8 & 50 & n/a & n/a \\
Console & 5 & 24 & 7 & 40 \\
Drop Frame & 5 & 12 & 3 & 5 \\
Breakpoints View & 3 & 13 & 2 & 110 \\
Unit View & 3 & 7 & n/a & n/a \\
\bottomrule
\end{tabular}}
\end{table}

\tabref{fig:study_features_used} shows the number of participants who used the various views and features of \visuflow and Eclipse, and the median time they spent on each feature. Due to technical difficulties, we were able to process the logs of only \empirical{14} participants.
As expected, the \emph{Java Editor} is the most commonly used feature. However, the \emph{Jimple View} was often used, showing that access to the intermediate representation is helpful when debugging static analysis. Other frequently used features include breakpoints, stepping, and variable inspection.
The \visuflow-exclusive features that were used the most are the \emph{Graph View} and the \emph{Results View} (\empirical{100}\% and \empirical{57.1}\% of participants, respectively).

Using \visuflow, participants spent \empirical{25.6}\% less time in the \emph{Java Editor}, and \empirical{44.4}\% less time stepping through code. Instead, they spent this time using the \emph{Graph View}, the \emph{Results View}, and the \emph{Variable Inspection View}. This shows that graph visualizations and access to the intermediate results of the analysis are desirable features for debugging.
Participants used the \emph{Breakpoints View} \empirical{88.2}\% less often in \visuflow compared to Eclipse. We attribute this to the special breakpoint features in \visuflow that allow users to step through both code bases simultaneously, sparing them the effort of writing conditional breakpoints in the \emph{Breakpoints View}.

The \emph{Unit View} was only used by \empirical{3} participants, all of whom are unfamiliar with Jimple. We believe that the \emph{Unit View} may be more popular for tasks requiring more knowledge about Jimple statements (e.g., writing an analysis rather than debugging it). However, we cannot verify this hypothesis through our study.

Using a $\mathcal{X}^2$ test of independence, we did not find a significant correlation \empirical{($p > 0.05$)} between the participants' background, their Net Promoter Scores, and the tool features they used the most.

\roundbox{
\RQO{6}{1}: Graphs, special breakpoints, and access to the intermediate representation and to the intermediate results are desirable features in a debugging environment for static analysis.
}

\subsubsection{\RQ{7}: Does \visuflow help identify and fix more errors compared to the standard debugging environment in Eclipse?}
\begin{table}[t]
\centering
\caption{Number of errors identified (I) and fixed (F) with Eclipse (E) and \visuflow (V) by each participant.}
\label{fig:study_errors}
\resizebox{\columnwidth}{!}{
\begin{tabular}{ l r r r r r r r r r r }
\toprule
 & \multicolumn{2}{ c }{\textbf{Task 1 (E)}} & \multicolumn{2}{ c }{\textbf{Task 2 (V)}} & & & \multicolumn{2}{ c }{\textbf{Task 1 (V)}} & \multicolumn{2}{ c }{\textbf{Task 2 (E)}} \\
 \cmidrule(lr){2-3} \cmidrule(lr){4-5} \cmidrule(lr){8-9} \cmidrule(lr){10-11}
 & \textbf{I} & \textbf{F} & \textbf{I} & \textbf{F} & & & \textbf{I} & \textbf{F} & \textbf{I} & \textbf{F} \\
\midrule
\PA{1} & 0 & 0 & 1 & 1 & & \PA{11} & 2 & 2 & 1 & 1 \\
\PA{2} & 0 & 0 & 1 & 1 & & \PA{12} & 1 & 0 & 2 & 1 \\
\PA{3} & 1 & 1 & 1 & 1 & & \PA{13} & 2 & 2 & 1 & 1 \\
\PA{4} & 1 & 0 & 1 & 1 & & \PA{14} & 2 & 1 & 0 & 0 \\
\PA{5} & 0 & 0 & 0 & 0 & & \PA{15} & 1 & 1 & 0 & 0 \\
\PA{6} & 3 & 3 & 3 & 3 & & \PA{16} & 1 & 1 & 2 & 1 \\
\PA{7} & 2 & 1 & 2 & 2 & & \PA{17} & 2 & 1 & 1 & 1 \\
\PA{8} & 2 & 1 & 0 & 0 & & \PA{18} & 2 & 1 & 1 & 1 \\
\PA{9} & 2 & 1 & 0 & 0 & & \PA{19} & 3 & 2 & 2 & 1 \\
\PA{10} & 1 & 1 & 2 & 2 & & \PA{20} & 1 & 0 & 0 & 0 \\
\midrule
\textbf{Sum} & 12 & 8 & 11 & 11 & & & 17 & 11 & 10 & 7 \\
\bottomrule
\end{tabular}}
\end{table}

\tabref{fig:study_errors} reports the number of errors identified and fixed by each participant. An error is \emph{identified} when a participant could explain why it occurred in the analysis code. \emph{Fixed} errors are identified errors that the participant fixed later. For Task~1, participants identified and fixed \empirical{1.4$\times$} more errors with \visuflow than with Eclipse. In particular, \empirical{17} errors were identified and \empirical{11} were fixed with \visuflow compared to \empirical{12} and \empirical{8} with Eclipse for that task. For Task~2, participants identified \empirical{1.1$\times$} and fixed \empirical{1.6$\times$} more errors when using \visuflow. 
Overall, more participants identified (\empirical{11}) and fixed (\empirical{10}) errors using \visuflow compared to Eclipse. Using Eclipse, only \empirical{4} and \empirical{3} participants identified and fixed more errors, respectively. The remaining participants found and fixed the same number of errors with both tools.
We do not compare the number of errors found by the same participant with different tools, because the two tasks were run on different, and thus incomparable, analyses.

A two-tailed Wilcoxon rank sum test does not show correlations between the tool used and the number of identified or fixed errors (\empirical{$p > 0.05$}). 
This is due to the high number of Eclipse users among our participants (\empirical{12/20}), causing a much lower learning curve for Eclipse.
Despite this factor, \empirical{7} of those \empirical{12} participants found and fixed more errors with \visuflow than with their original debugging environment. Moreover, we found no significant correlations between the number of errors identified and fixed, and participant background (coding environment, seniority, knowledge of Soot, Eclipse, and data-flow analysis).

\roundbox{
\RQO{7}{1}: Using \visuflow helped participants identify \empirical{25\%} and fix \empirical{50\%} more errors compared to using Eclipse.
}

\subsubsection{\RQ{8}: Does \visuflow help understand and debug analyses better than other debugging environments?}
\label{subsec:study_usefulness}

After performing the tasks, participants filled out a comparative questionnaire to assess the perceived usefulness of various debugging environments. They rated \visuflow, Eclipse, and their own debugging environment through a Net Promoter Score (NPS)~\cite{nps}. Additionally, participants evaluated how \visuflow and Eclipse helped them perform the required tasks, and identified their preferred features in both debugging environments.

Overall, \visuflow was positively received. In the NPS questions, the \empirical{20} participants rated their likelihood of recommending a debugging environment over another one to a friend, for a task similar to the ones that they performed in the study. \visuflow has a mean NPS score of \empirical{9.1} out of 10 (standard deviation \empirical{$\sigma = 1.1$}) compared to Eclipse, and \empirical{8.3} (\empirical{$\sigma = 1.7$}) compared to the participant's own debugging environment. Eclipse has a mean score of \empirical{1.4} (\empirical{$\sigma = 1.6$}) compared to \visuflow, and \empirical{3.4} (\empirical{$\sigma = 3.3$}) compared to the participant's own debugging environment.

We then asked participants which debugging environment made it easier for them to find/fix the errors and understand the static analysis code. \empirical{All} participants answered that identifying errors was easier with \visuflow (\quot{It is pretty obvious that that's what static analysis needs.}). \empirical{Sixteen} participants found it easier to fix errors with \visuflow; the other \empirical{4} participants answered that both debugging environments made it equally easy. \empirical{Seventeen} participants wrote that \visuflow helped them understand the analysis code better (\quot{What I was looking for in the first coding environment [Eclipse] was given to me by the second one \visuflowbr}), while \empirical{1} participant preferred Eclipse, and \empirical{2} participants remained neutral. To our surprise, the \empirical{12} participants who were already familiar with Eclipse still preferred \visuflow, showing that \visuflow is better suited than traditional debugging tools for debugging static analysis.

When asked what they would use both debugging environments for, \empirical{16} participants wrote they would use \visuflow to write and debug static analysis (\quot{[I would use \visuflow for] visualising an analysis and finding unexpected values included or excluded from expected results}). \empirical{Eleven} participants found Eclipse more useful for \quot{standard software development} or \quot{general Java programming}.

Participants were asked to write in free-text which features of \visuflow and Eclipse they would like to have in their own debugging environments. \empirical{Three} participants liked Eclipse's integrated debugger, which echoes our survey findings (\tabref{fig:survey_features_liked}). We received more requests for features that \visuflow provides. In particular, \empirical{10} participants asked for the \emph{Graph View} (\quot{visualising for provenance was useful}). \empirical{Seven} participants asked for visualizing intermediate results (\quot{\visuflowbr is useful, because I get the abstract view of the situation, what's happening inside. Before [with the other coding environment], you have to [go through all] the variables.}). \empirical{Five} participants asked for the special breakpoints (\quot{\visuflowbr is more comfortable; you can set Jimple breakpoints. It is clearly better.}). \empirical{Three} participants asked for the synchronization between multiple views (\quot{I think \visuflowbr is helpful because of the linkage between the Java code, the Jimple code and the graphic visualization: all that I had to keep in my mind [earlier]}). The \emph{Jimple View} and the \emph{Unit Inspection View} were only mentioned \empirical{once}. \empirical{Two} novice Soot users wrote that they wanted \quot{All of them}.
The features that participants find useful the most confirm our survey findings (\RQO{5}{3}), and match the participants' behaviour (\RQO{6}{1}).

Novice analysis writers noted a gentler learning curve when using \visuflow. A participant said: \quot{I think this approach of debugging in the CFG is easier to learn for starting with taint analysis}, and another one noted: \quot{For someone who doesn't do this style of debugging analysis code at all, it kind of surprised me how quickly I was able to track down bugs for a bunch of code that I don't understand.}

\roundbox{
\RQO{8}{1}: Participants find \visuflow more useful than Eclipse and than their own tools to debug static analysis code. In the questionnaire and interviews, participants confirm that the features identified as most important in our survey allow novice and expert analysis writers to more easily understand and fix bugs in analysis code.
}
\section{Threats to Validity}
\label{sec:limitations}

We noticed that \Q{11} of the survey was misinterpreted by a few participants, because their answers do not match the explanation given in \Q{12}. For example, a participant wrote in \Q{12} that \quot{debugging SA [static analysis] is still a bit harder [than application code]}, and gave \Q{11} a score of 7 (scale from 1 to 10), denoting that application code is harder to debug than analysis code. In such clear cases, we reversed the score (in this example, the new score is 4) and reported the new score. We reversed only \empirical{12} scores out of \empirical{103} responses.

We conducted the user study in a controlled environment rather than in a development setting. Therefore, the setting of \empirical{20} participants, \empirical{20} minutes per task, and the analyses that were used in the user study are not quite representative of realistic uses. In practice, users would have more time to investigate much more complex analyses. Given the time limits, we had to simplify the analyses. To make them as realistic as possible, we based them on taint analyses written by experienced students in our graduate course. We then introduced typical errors made by those students. The resulting analyses are \empirical{\textasciitilde 300}~LOC long. We verified with our pilot participants that the tasks could be achieved in the time limits. To avoid further external threats to validity, we recruited participants from different backgrounds: academia, industry, students, and professionals. \visuflow is built on top of Eclipse and Soot, which are well-established both in industry and academia.
It would, however, be interesting to conduct a future study with more industry participants in real-life conditions.

The times we report in \tabref{fig:study_features_used} represent the mouse focus time on different views. These times are not exact, since participants attention may be divided between multiple views while the mouse can focus on only one of them.
We argue that, for our user study, participants mainly used the mouse to navigate between views. In the absence of an eye-tracking device, our measurements approximate real user data. Averaged over all users, the relative difference between the times spent in each view is still a reliable metric.

\section{Discussion and Future Work}

Our survey collected extensive data, not only about debugging features for static analysis, but also about debugging features for general application code, motivations for writing static analysis (\Q{5}), types of analysis written by participants (\Q{2}, \Q{7}, \Q{8}), detailed analysis examples (\Q{10}), reasons why participants debug static analysis (\Q{14}), and why participants use a particular debugging environment (\Q{30}). Due to space limitation, we did not report all data, but it is available online for further use~\cite{visuflow}.

We designed the current \visuflow prototype as a proof of concept to confirm the usefulness of some of the features identified in our survey. However, \visuflow, in its current version, does not scale to more complex analyses that need to handle larger graphs that would not be easily understandable. For such analyses, \visuflow has a higher latency while waiting for the analysis to terminate. We plan to address both issues in future work. 
Furthermore, it would be interesting to investigate and integrate more of the debugging features found in the survey (e.g., omniscient debugging and quick updates) in debugging tools like \visuflow.
\visuflow is available online~\cite{visuflow}, and we encourage contributions by other researchers and practitioners.

\section{Related Work}

Debugging static analysis has not been a major topic in the software engineering community. In this section, we discuss prior work on visualizing static analysis information, existing debugging tools and techniques, and the usability of static analysis tools.

\subsection{Debugging Static Analysis}
We are not aware of any tool that is tailored to address issues specific to debugging static analysis. In past work, Andreasen et al.~\cite{andreasen2017} suggest to employ soundness testing, delta debugging, and blended analysis to debug static analysis. By means of a few examples, they discuss how the combination of these techniques (both pairwise and all three of them) have helped them locate and fix bugs in their static analyzer TAJS. Other tools also provide a subset of the features in \visuflow, especially in terms of visualization of information and data flows. For example, Atlas~\cite{Deering:2014} visualizes data-flow paths based on the abstract syntax tree (AST) of a given program. To improve user understanding and evaluating error reports, Phang et al.~\cite{Khoo:2008} present a tool that visualizes program paths to help the user track where an error originates from. Unlike \visuflow, none of these tools enable static analysis writers to debug their own analyses, but are rather tailored to the users of static analysis tools (e.g., code developers), and therefore focus more on visualization features than debugging features.

\subsection{Standard Debugging Tools}
Most programming languages' runtime environments are shipped with debuggers provided by the language maintainers (e.g., GDB~\cite{gdb}). Many IDEs, such as Eclipse~\cite{eclipse} and IntelliJ~\cite{intellij}, integrate debugging functionalities for major programming languages natively in their tool sets. Since \visuflow is integrated into Eclipse, it uses all available features such as breakpoints and stepping. As our survey and user study have shown, such tools are designed for general application code, and do not have specific support for static analysis. There exist more complex debugging techniques such as delta debugging~\cite{ZellerH02}, omniscient debugging~\cite{Lewis03}, and interrogative debugging~\cite{KoM04}. However, these techniques are not integrated into the most commonly used tools such as Eclipse. 
 
\subsection{Usability of Static Analysis Tools}
To our knowledge, this paper presents the first large-scale survey of static analysis developers. Most of the prior surveys of developers are targeted towards static analysis users instead of writers. Ayewah et al.~\cite{Ayewah2008} present a survey of FindBugs~\cite{findbugs} users to determine their usage habits and how they deal with the displayed warnings. The authors conclude that static analysis tools have been widely adopted by their participants, but are not used regularly, and without customization. Christakis and Bird~\cite{Christakis:2016} asked 375 developers within Microsoft about their attitudes towards static analysis tools. The features that  participants deemed most important include: better usability, better responsiveness, and pre-configured prioritization of security and best-practice aspects when it comes to error reporting. Johnson et al.~\cite{Johnson:2016} investigated the warning and error reports of the static analyzer FindBugs, the Eclipse Java Compiler, and the code-coverage tool Eclemma~\cite{eclemma}. Nguyen Quang Do et al.~\cite{cheetah} evaluated the impact of their Just-in-Time static analysis on the workflow of developers who use such tools to detect errors in their code. Phang et al.~\cite{Khoo:2008} also tested the program-path visualization tool discussed above through a user study. However, the goal of those surveys and user studies is to assess the usability of static analyzers from the end-user perspective. This paper, on the other hand, aims at collecting requirements for a debugging tool for static analysis writers, and assess the usability of such a tool.

\section{Conclusion}
Writing and debugging static analysis is a difficult task. We surveyed \empirical{115} static analysis writers from different backgrounds and show that current debugging tools are not always sufficient to properly support static analysis writers.  In this paper, we report the main causes of bugs in static analysis, show the major tool features used by analysis writers to debug their analyses, discuss their limitations, and identify features that would best support debugging static analysis.

We present \visuflow, a debugging environment designed specifically for debugging static analysis, including some of the features we identified in our survey. In a comparative user study between \visuflow and Eclipse, we empirically show that \visuflow enables analysis writers to debug static analysis more efficiently. \visuflow was well received by analysis writers, confirming our survey's findings, and validating the usefulness of its debugging features.

The full list of debugging features presented in this paper can be used to design better tool support for debugging static analysis and make it easier for analysis writers to secure application code.

\section*{Acknowledgements}
We thank the participants of our surveys and user studies for their invaluable input. We would also like to thank Henrik Niehaus, Shashank Basavapatna Subramanya, Kaarthik Rao Bekal Radhakrishna, Zafar Habeeb Syed, Nishitha Shivegowda, Yannick Kouotang Signe, and Ram Muthiah Bose Muthian for their work on the implementation of \visuflow.
This research was supported by a Fraunhofer Attract grant as well as the Heinz Nixdorf Foundation. This work has also been partially funded by the DFG as part of project~E1 within the CRC 1119 CROSSING, and was supported by the Natural Sciences and Engineering Research Council of Canada.

\bibliographystyle{ACM-Reference-Format}
\bibliography{references}

\end{document}